\documentclass[twocolumn,nofootinbib,amsmath,amssymb,preprintnumbers]{revtex4}

\usepackage{graphicx}
\usepackage{slashbox}
\usepackage{color}
\usepackage{epstopdf}

\newcommand{\squishlist}{
   \begin{list}{$-$}
    { \setlength{\itemsep}{0pt}      \setlength{\parsep}{3pt}
      \setlength{\topsep}{4pt}       \setlength{\partopsep}{0pt}
      \setlength{\leftmargin}{1.5em} \setlength{\labelwidth}{1em}
      \setlength{\labelsep}{0.7em} } }

\newcommand{\squishlisttwo}{
   \begin{list}{-}
    { \setlength{\itemsep}{0pt}    \setlength{\parsep}{0pt}
      \setlength{\topsep}{0pt}     \setlength{\partopsep}{0pt}
      \setlength{\leftmargin}{2em} \setlength{\labelwidth}{1.5em}
      \setlength{\labelsep}{0.6em} } }

\newcommand{\squishend}{
    \end{list}  }

\newcommand{\tbktitle}[1]{``#1''}     % book title
\newcommand{\tISBN}[1]{#1} % ISBN number

\newcommand{\bea}{\begin{eqnarray}}
\newcommand{\eea}{\end{eqnarray}}
\newcommand{\beq}{\begin{equation}}
\newcommand{\eeq}{\end{equation}}

\begin{document}
\title{Self-Organized Criticality, effective dynamics and the universality class of the Deterministic Lattice Gas}

\author{Andrea Giometto$^{1,2}$}
\email{andrea.giometto10@imperial.ac.uk}
\author{Henrik Jeldtoft Jensen$^{3}$}
\email{h.jensen@imperial.ac.uk}
\affiliation{$^1$ 
Blackett Laboratory, Department of Physics and Complexity \& Networks Group, Imperial College London, London, SW7 2AZ, UK}
\affiliation{$^2$ Dipartimento di Fisica G. Galilei,  Universit\`a di Padova, Via Marzolo 8, I-35151 Padova, Italy}
\affiliation{$^3$ Complexity \& Networks Group and Department of Mathematics , Imperial College London, London, SW7 2AZ, UK}
 
\begin{abstract}
We show that the Deterministic Lattice Gas (DLG) (Phys. Rev. Lett. {\bf 64}, 3103 (1990)) model of Self-Organized Criticality (SOC) despite of its deterministic micro dynamics belongs to the Manna universality class of absorbing state phase transitions. This finding is consistent with our observation that an effective stochastic term is generated in the DLG  at large length scales, whereby the macro dynamics of the DLG appears closer to the other stochastic SOC models of the Manna class.  
\end{abstract}

\maketitle

It is well known that Self-Organized Criticality (SOC) was introduced by Bak, Tang and Wiesenfeld (BTW) as an attempt to explain the widespread occurrence of $1/f$ temporal fluctuations and fractal spatial structure\cite{Bak_1987,HJJ_book}.  It was, however, soon realized that the model used by BTW did not contain a $1/f$ spectrum\cite{HJJ_0,HJJ_book}.  The Deterministic Lattice Gas (DLG) was introduced in an effort to find another deterministic model, which could be used as a proof of existence for the SOC scenario. The model being deterministic is difficult to analyze  analytically and one has to rely on a combination of simulations and non-rigorously 
justified effective analytic investigations\cite{Grinstein1992}. Numerical simulations found that the density fluctuations in the DLG did exhibit $1/f$ fluctuations (and that dissipation did take place on a fractal). This was found to be consistent with analysis of (non-linear) diffusion equations for which one assumes the absence of any bulk noise term\cite{Grinstein1992}. 

Until now it has been unclear how the DLG related to the universality classes of SOC\cite{Lubeck_5,Pruessner_book} and how the existence of an absorbing state at low density influences the behavior of the model. In this letter we first present simulations which demonstrate that the density fluctuations at elevated densities for large systems have a power spectrum $S(f)\propto 1/f^\mu$ with $\mu = 3/2$, which is consistent with a description in terms of diffusion equation with a  bulk noise term. Next we present compelling evidence that the transition to an absorbing state is of the Manna universality class and use scaling relations to determine the value of the power spectrum exponent $\mu$ for densities near the absorbing state phase transition.

{\em Model --} In the DLG particles interact through a nearest-neighbor repulsive central unit force and double occupancy  is not permitted.  All particles are updated simultaneously by moving the particles {\em deterministically} to neighbor sites   according to the vector sum of the forces they are subject to. If two particles want to move to the same site, the particle subject to the strongest force is moved, while in the case of equal forces no particle is displaced. Periodic boundary conditions are considered and we consider the model in dimension $d=2$.

The number of particles $N(t)$ in a sub-volume of the lattice exhibits interesting temporal fluctuations. We determine the power spectrum $S(f)$ of $N(t)$ from the square of the absolute value of the Fourier transform. Successive time sequences are averaged in order to achieve sufficient statistics:
\beq
S(f) = \frac{1}{2 \pi T} \left\langle \left\vert \sum_{t=1}^{T} N(t) e^{-i 2 \pi f t} \right\vert^2 \right\rangle
\label{S(f)_measured}
\eeq
The angular brackets denote averaging over many different time series.

\begin{figure}[!hbt]
\begin{center}
\includegraphics[width=6cm,angle=-90]{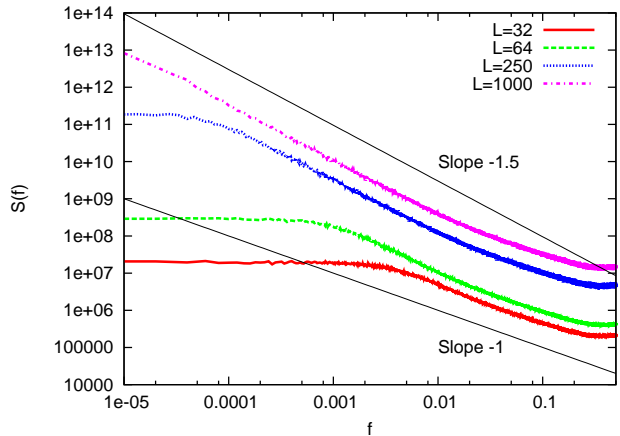}
\caption{Scaling behavior of the spectrum $S(f) \propto f^{-\mu}$ of the total number of particles $N(t)$ in the DLG for increasing linear sizes of the lattice $L$. A crossover from $\mu = 1$ for small $L$ to $\mu = 1.5$ for large $L$ is observed. Particle density $\rho=0.5$.}
\label{S_size_depen}
\end{center}
\end{figure}

The spectrum $S(f)$ was shown\cite{HJJ_5} to satisfy $S(f) \sim 1/f^\mu$ at the particle density $\rho = 0.3$, while previous works showed that the same result is obtained in a wide range of densities with a drive at the boundary\cite{HJJ_1}. In those papers the maximum linear system size considered is $L=128$, we will see that different behaviors are observed for larger systems. In Fig. \ref{S_size_depen} we show how the effective exponent $\mu$ changes from $\mu=1$ to $\mu=3/2$ as $L$ increases. It is interesting to relate this change in the temporal fluctuations to the appropriate macroscopic equation of motion for the particle density $\rho({\bf r},t)$.

Simulations show that although particles move deterministically in the DLG model they do behave like random walkers in the sense that their square displacement is linear in time. This suggests that $\rho({\bf r},t)$ evolves according to a diffusion equation and integrating the diffusion equation allows us to extract the temporal evolution of $N(t)=\int_V d{\bf r} \rho({\bf r},t)$, where $V$ denotes the measuring sub-volume. Assume that $\rho({\bf r},t)$ satisfies the diffusion
\beq
\frac{\partial \rho}{\partial t} = D\nabla^2\rho + \xi.
\label{diffuse}
\eeq
It has been shown\cite{Grinstein1992,HJJ_book} that the power spectrum of $N(t)$ depends on the details of the diffusion equation. If the equation is driven by white noise at the boundary of $V$ and the noise term $\xi$ is absent the power spectrum exponent derived from Eq. (\ref{diffuse}) is $\mu=1$. If in contrast a conservative bulk noise term is included in Eq. (\ref{diffuse}) one obtains instead $\mu=3/2$. Including non-linearities in Eq. (\ref{diffuse}) will not influence the behavior of $\mu$, see \cite{Grinstein1992}. It is this observation that allows us to conclude that a conservative noise term is generated in the macroscopic Langevin description when the sub-volume $V$ becomes sufficiently large.\\

\begin{figure}[!hbt]
\begin{center}
\includegraphics[width=6cm,angle=-90]{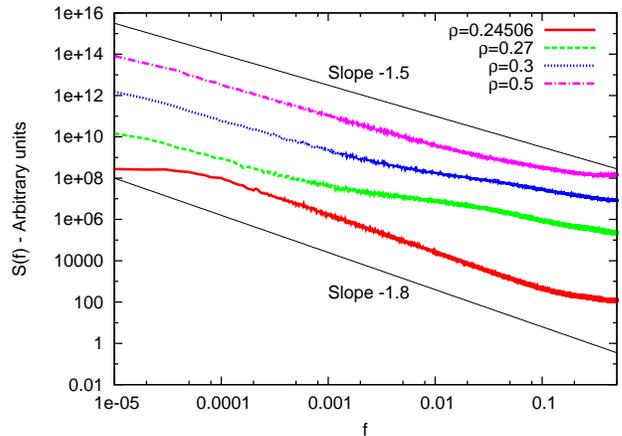}
\caption{Scaling behavior of the spectrum $S(f) \propto f^{-\mu}$ of the total number of particles $N(t)$ in the DLG for different particle densities $\rho$. A crossover from $\mu \simeq 1.8$ at $\rho \simeq \rho_c=0.245$ to $\mu = 1.5$ at $\rho \gg \rho_c$ is observed. $S(f)$ has been multiplied by different constants for different densities $\rho$ to visualize the scaling exponents properly. Lattice size $L=1000$.}
\label{spectra_pdlg}
\end{center}
\end{figure}

{\em $S(f)$ scaling for large lattices --} We now consider the DLG for much larger lattices than previously studied and our aim is to understand the fluctuations spectrum as function of density.  We plot $S(f)$ in Figure \ref{spectra_pdlg} and find that the value of $\mu$  depends on the density:
\begin{equation}
 S(f) \sim
 \begin{cases}
 f^{-1.5} & \rho \geq 0.3 \\
 f^{-1.8} & \rho \simeq \rho_c = 0.245
 \end{cases}
 \end{equation}
where $\rho_c$ is the critical density of the absorbing phase transition (APT in the following).

The observed exponent at high densities is the same of a gas of random walkers \cite{HJJ_2}, while  at low densities the power spectrum scaling exponent is determined by the critical properties of the DLG at the APT, as we shall discuss in the following.\\

{\em Absorbing Phase Transition --}
At very low densities the DLG enters configurations in which all particles are far away from each other and, due to the short range interactions, the particles become unable to move, i.e. the dynamics is frozen.  Near the transition 
 the density of active particles $\rho_a$ behaves according to $\rho_a(\delta\rho) \sim  \delta\rho^{\beta}$ for $ \rho > \rho_c$ and $\rho_a(\delta\rho)  =0$ for $ \rho < \rho_c$
  with $\delta\rho=\rho-\rho_c$. We determine the critical density $\rho_c$ from a log-log plot of $\rho_a$ versus $\delta \rho$ and obtain  $\rho_c=0.24500(2)$.  A regression analysis yields the value of the order parameter exponent $\beta=0.634(2)$.
 
We find that close to the critical point the fluctuations of the order parameter
 $\Delta \rho_a = L^2 \left[ \langle \rho_a^2 \rangle - \langle \rho_a \rangle^2 \right]\sim\delta\rho^{-\gamma'}$ with $\gamma'=0.40(2)$.

To circumvent finite size effects we make use of the implementation of the external field for APTs   developed in \cite{Lubeck_2}: at each time step, after performing the DLG update, we choose randomly $h L^2$ particles on the lattice and move each of them to one of its empty neighbors.  The external field prevents the system from falling into an absorbing state. We assume that the order parameter and its fluctuations satisfy $
\rho_a (\delta \rho, h) = \lambda^{-\beta} \tilde R (\delta\rho \lambda, h \lambda^{\sigma})
$ and $\Delta \rho_a (\delta \rho, h) = \lambda^{\gamma'} \tilde D (\delta\rho \lambda, h \lambda^{\sigma})$,
where $\lambda>0$. From this scaling ansatz we conclude that $ \rho_a(\delta\rho=0,h)\sim h^{\beta/\sigma}$ and $\Delta\rho_a(\delta\rho=0,h)\sim h^{-\gamma'/\sigma}$. Our simulations lead to  $\sigma=2.19(1)$ \cite{footnote} and  $\gamma'=0.37(1)$.

\begin{figure}[!hbt]
\begin{center}
\includegraphics[width=6cm,angle=-90]{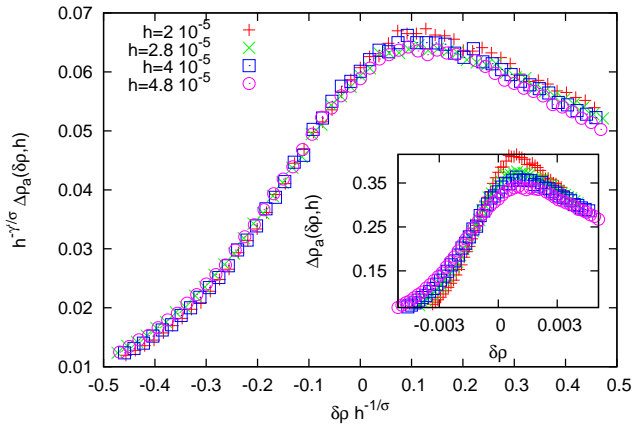}
\caption{Data collapse of $\Delta\rho_a$. In the inset data before the collapse are shown. Lattice linear size $L=1000$.}
\label{collapse_fluct}
\end{center}
\end{figure}

Data collapses can be produced choosing $\lambda=h^{-1/\sigma}$: see Fig.  \ref{collapse_fluct} for $\Delta\rho$ and \cite{Giometto_to_be_published} for $\rho_a$ and more details.
%
%\begin{figure}[!hbt]
%\begin{center}
%\includegraphics[width=6cm,angle=-90]{figures/scaling/collapse_rhoa}
%\includegraphics[width=\columnwidth]{figures/scaling/collapse_rhoa}
%\includegraphics[width=\columnwidth]{figures/scaling/collapse_rhoa}
%\caption{Data collapse $\rho_a$. In the inset data before the collapse are shown. Lattice linear size $L=1000$.}
%\label{collapse_rhoa}
%\end{center}
%\end{figure}
%
%

{\em Finite Size Scaling --} We assume the following finite size scaling form $\rho_a (\delta \rho, h, L) = \lambda^{-\beta} \tilde R_{pbc} (\delta\rho \lambda, h \lambda^{\sigma}, L \lambda^{-\nu_{\perp}})$ and $\Delta \rho_a (\delta \rho, h, L) = \lambda^{\gamma'} \tilde D_{pbc} (\delta\rho \lambda, h \lambda^{\sigma}, L \lambda^{-\nu_{\perp}})$, where the exponent $\nu_{\perp}$ describes the divergence of the spatial correlation length, i.e., $\xi_{\perp} \propto \delta \rho^{-\nu_{\perp}}$. The universal scaling functions depend on the particular choice of boundary conditions, although in the thermodynamic limit $\tilde R_{pbc}(x,y,\infty)=\tilde R(x,y)$ and $\tilde D_{pbc}(x,y,\infty)=\tilde D(x,y)$.

\begin{figure}[!hbt]
\begin{center}
\includegraphics[width=6cm,angle=-90]{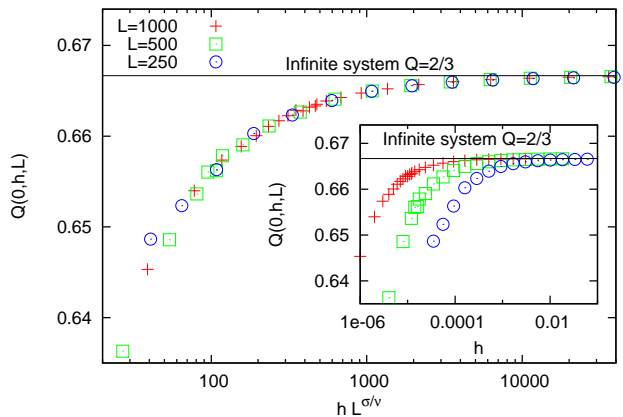}
\caption{(lin-log) Data collapse of $ Q(0, h, L) $  for the determination of $\nu_{\perp}$. In the inset data before the collapse are shown.}
\label{nu_collapse}
\end{center}
\end{figure}

Following \cite{Lubeck_5} we consider the fourth order cumulant $Q$, which is defined as: $Q = 1 - \frac{\langle \rho_a^4 \rangle}{3 \langle \rho_a^2 \rangle^2}$. For non-vanishing order-parameter the cumulant tends to $Q=2/3$ in the thermodynamic limit. One expects that the cumulant obeys the scaling form: $Q(\delta \rho, h, L) = \tilde Q_{pbc} (\delta \rho \lambda, h \lambda^{\sigma}, L \lambda^{-\nu_{\perp}})$.
Choosing $\lambda=L^{1/{\nu_{\perp}}}$  at $\delta \rho = 0$ we obtain the following equation:
\begin{equation}
Q(0, h, L) = \tilde Q_{pbc} (0, h L^{\sigma/{\nu_{\perp}}}, 1)
\label{nu}
\end{equation}
which enables us to determine $\nu_{\perp}$ through a data collapse by plotting $Q(0,h,l)$ against ${hL^{\sigma/\nu_{\perp}}}$ as in Fig. \ref{nu_collapse}. Best results are obtained for $\nu_{\perp}=0.83(5)$.

{\em Dynamical scaling -- } Starting simulations of the DLG from a random distribution of particles above the critical point $\rho_c$ the density of active sites $\rho_a$ decreases in time and tends to its steady state value. At the critical point $\rho=\rho_c$ the order parameter decays algebraically as $\rho_a(\delta \rho=0, h=0,t) \sim t^{-\alpha}$. Simulating a lattice of linear size $L=4000$ we obtain $\alpha=0.41(1)$. From the scaling ansatz $\rho_a(L,t) = L^{-\alpha z} \tilde R_{pbc}'(t L^{-z},1)$, with $z = \nu_{\parallel}/\nu_{\perp}$.  The best data collapse is obtained for $z=1.5(1)$. For more details see \cite{Giometto_to_be_published}.

{\em DLG universality class --} We are now able to address the question concerning which universality class the DLG belongs to. In table \ref{DLG_Manna} we compare the measured values of the DLG critical exponents with those of the Manna universality class\cite{Lubeck_book}, showing that they are compatible with each other - for further details see \cite{Giometto_to_be_published}.
\begin{table}[!hbt]
\begin{center}
\begin{tabular}{|c|c|c|c|c|}	\hline
	&	$\beta$		&	$\nu_\perp$	&	$\nu_\parallel$	&	$\sigma$\\	\hline
Manna	&	0.639(9)	&	0.799(14)	&	1.225(29)	&	2.229(32)	\\
DLG		&	0.634(2)	&	0.83(5)	&	1.2(1)	&	2.19(1)\\	\hline	\hline
	&	$\gamma'$	&	$\gamma$	&	$\alpha$	&	$z$	\\	\hline
Manna	&	0.367(19)	&	1.590(33)		&	0.419(15)		&	1.533(24)	\\
DLG 		&	0.37(1)	&	1.54(1)		&	0.41(1)		&	1.5(1) \\ \hline
\end{tabular}
\caption{The measured critical exponents for the DLG and the corresponding critical exponents for the Manna universality class in $d=2$ \cite{Lubeck_book}.}
\label{DLG_Manna}
\end{center}
\end{table}
The shape of the universal scaling functions is remarkably similar to those that can be found in the literature for the Manna universality class. This is seen e.g. by comparing our Fig. \ref{collapse_fluct} with Fig. 4 in \cite{Lubeck_2}, see also \cite{Lubeck_book}. From the scaling exponents and the scaling functions we find compelling evidence that the DLG belongs to the Manna universality class.
 
{\em Power spectrum at criticality --} Finally we  relate the exponent of the power spectrum at low density to the scaling properties of the DLG at the critical point $\rho_c$ of the APT\cite{Lauritsen}. The scaling behavior of the correlation function $C_a(\mathbf{r},t)=\langle \rho_a(\mathbf{r},t) \rho_a(0,0) \rangle - \langle \rho_a(0,0)^2 \rangle$ at the critical density $\rho=\rho_c$ determines the power law exponent in the scaling of the spectrum $S_a(f)$ of the total number of \textit{active} particles $N_a(t)$ in the lattice. Assume
\begin{equation}
C_a(\mathbf{r},t) = \lambda^{-\eta} \tilde C_a (\lambda \rho, \lambda^{-\nu_\parallel} t, \lambda^{-\nu_{\perp}} r)
\end{equation}
 In stationary directed percolation processes above criticality it has been observed\cite{Lubeck_book} that the correlation function $C_a(\mathbf{r},t)$ at $t=0$ first decays in space algebraically as $r^{-\beta/\nu_\perp}$, until it saturates at a constant value at $r > \xi_\perp$. In the saturated regime the two sites become uncorrelated so that this value is just equal to the squared stationary density of active sites $\rho_a^2$. One can use this fact to extract the scaling exponent for $S_a(f)$, making use of the Wiener-Khinchin theorem:
\begin{align*}
C_a(t) &= \langle N_a(t) N_a(0) \rangle - \langle N_a(0)^2 \rangle \\ 
&= \left\langle \int_V d^2\mathbf{r} \int_V d^2\mathbf{r'} \rho_a(\mathbf{r},t) \rho_a(\mathbf{r'},0) \right\rangle + \dots \\
%&= \left\langle V \int_V d^2\mathbf{r} \ \rho_a(\mathbf{r},t)\rho_a(0,0) \right\rangle + \dots \ \\ 
&= \ \int_V d^2\mathbf{r} \ C_a(\mathbf{r},t) + \dots %\sim \int_V dr \ r \ r^{-\beta/\nu_\perp} \tilde C_a(t/r^z) \ \\
 \sim \ t^{1/z(2-\beta/\nu_\perp)} 
\end{align*}
so that:
\begin{equation*}
S_a(f) = \frac1{2\pi }\int dt \ C_a(t) \ e^{-i 2 \pi f t} \sim f^{-1-\frac{1}{z}\left(2-\frac{\beta}{\nu_\perp}\right)} \sim f^{-\mu}
\end{equation*}
with $\mu = 1+\frac{1}{z}\left(2-\frac{\beta}{\nu_\perp}\right)$. With our estimates of the critical exponents for the DLG we find $\mu=1.82(6)$, while with the best estimates of the Manna universality class exponents \cite{Lubeck_book} one finds $\mu=1.78(2)$.
\begin{figure}[!hbt]
\begin{center}
\includegraphics[width=6cm,angle=-90]{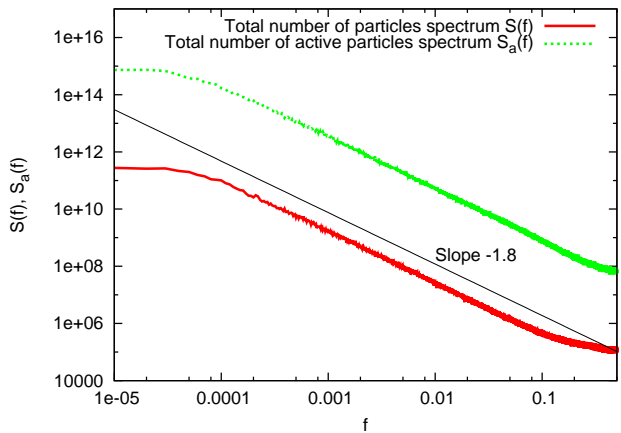}
\caption{DLG: Scaling behavior of the spectrum $S(f) \propto f^{-\mu}$ of the total number of particles in the box $N(t)$ and of the spectrum $S_a(f)$ of the total number of active particles near the critical density $\rho=0.24506 \simeq \rho_c$. The spectra have been multiplied by arbitrary factors to visualize the scaling exponent properly. Lattice size $L=1000$.}
\label{spectra_pdlg_critical}
\end{center}
\end{figure}
At low densities we expect that fluctuations in the total number of particles $N(t)$ in the lattice are triggered by active particles, therefore we expect the two spectra $S(f)=\frac1{2\pi}\int dt \ N(t) \ e^{-i 2 \pi f t}$ and $S_a(f) =\frac1{2\pi}\int dt \ N_a(t) \ e^{-i 2 \pi f t}$ to show the same scaling behavior at $\rho \simeq \rho_c$, as it is confirmed by simulations (see Fig. \ref{spectra_pdlg_critical}).
Computing the power spectrum of the total number of particles in the box at $\rho=0.24506 \simeq \rho_c$ we observe a power spectrum compatible with the predictions, confirming that fluctuations in the total number of particles in the sub-volume $V$ are determined by the scaling behavior of the correlation function at criticality.

{\em Summary -- } We have investigated the scaling properties of the power spectrum $S(f)$ of the total number of particles in a sub-volume of the lattice for the DLG: our simulations show that the DLG is not characterized by $1/f$ fluctuations as had been previously observed, but present a much more variegated picture. At high densities the power spectrum is the same as observed in a gas of random walkers, while at low densities the spectrum scales as $S(f) \sim 1/f^{1.8}$ and we have shown that this exponent is determined by the decay of the density-density  correlation function near criticality.

We have shown that the deterministic lattice gas is in the same universality class as the Manna model\cite{Manna}. To our knowledge this is the first example of a completely deterministic and non-chaotic system in this universality class. Perhaps this is not too surprising given that we found that a stochastic bulk noise is generated in the effective Langevin description as the system size is increased. From this point of view the DLG becomes similar to the other members of the Manna class, which all involved a stochastic element in their microscopic dynamics. Examples of models in the Manna class includes the Oslo model\cite{Oslo_Model} and the Manna model\cite{Manna}. In the Oslo model the local threshold for relaxation is updated stochastically at every relaxation and in the Manna model particles move to stochastically chosen neighbor sites.

%%%%%%%%%%%%%%%%%%%%%%%%%%%%%%%%%%%%%%%%%%%%%%%%%%%%

%%%%%%%%


\begin{thebibliography}{16}
\bibitem{Bak_1987}
	P. Bak, C. Tang, K. Wiesenfeld,
%	\tarttitle{Self-Organized Criticality: An Explanation of $1/f$ Noise}
	in {Physical Review Letters} 59, 381:384 (1987)

\bibitem{HJJ_book}
	H.J. Jensen,
	\tbktitle{Self-Organized Criticality. Emergent Complex Behavior in Physical and Biological Systems},
	Cambridge Lecture Notes in Physics, ISBN \tISBN{0-521-48371-9} (1998)	

\bibitem{HJJ_0}
	H.J. Jensen, K. Christensen, H.C. Fogedby,
%	\tarttitle{$1/f$ noise, distribution of lifetimes, and a pile of sand}
	in {Physical Review B} 40, 7425:7427 (1989)

\bibitem{Grinstein1992}
	G. Grinstein, T Hwa, H.J. Jensen,
%	\tarttitle{$1/f^{\alpha}$ noise in dissipative transport}
	in {Physical Review A} 45, R559:R562 (1992)

\bibitem{Lubeck_5}
	S. L\"ubeck, P.C. Heger,
%	\tarttitle{Universal finite-size scaling behavior and universal dynamical scaling behavior of absorbing phase transitions with a conserved field}
	in {Physical Review E} 68, 056102 (2003)

\bibitem{Pruessner_book}
	Gunner Pruessner,
	\tbktitle{Self-Organized Criticality. Theory, Models and Charaterisation},
	Cambridge University Press, ISBN \tISBN{9780521853354} (2011)	
	
\bibitem{HJJ_5}
	T. Fiig, H.J. Jensen,
%	\tarttitle{Diffusive Description of Lattice Gas Models}
	in {Journal of Statistical Physics} 71, 653:682 (1993)
	
\bibitem{HJJ_1}
	H.J. Jensen,
%	\tarttitle{Lattice Gas as a Model of $1/f$ Noise}
	in {Physical Review Letters} 64, 3103:3106 (1990)

\bibitem{HJJ_2}
	J.V. Andersen, H.J. Jensen, O.G. Mouritsen,
%	\tarttitle{Crossover in the power spectrum of a driven diffusive lattice-gas model}
	in {Physical Review B} 44, 439:442 (1991)

\bibitem{Lubeck_2}
	S. L\"ubeck,
%	\tarttitle{Scaling behavior of the order parameter and its conjugated field in an absorbing phase transition around the upper critical dimension} 
	in {Physical Review E} 65, 046150 (2002)

\bibitem{footnote}
The error on $\beta/\sigma$ is computed as the dispersion on its value when measured at $\rho_{c,1}=0.24498$ and $\rho_{c,2}=0.24502$ (respectively the lower and upper bound on our determination of the critical density $\rho_c=0.24500(2)$). Errors on other quantities are estimated in the same way, when possible.
 	
\bibitem{Giometto_to_be_published}
         A. Giometto and H.J. Jensen, to be published.

\bibitem{Lubeck_book}
	M. Henkel, H. Hinrichsen, S. L\"ubeck,
	\tbktitle{Non-Equilibrium Phase Transitions. Volume I Absorbing Phase Transitions},
	Springer, ISBN \tISBN{978-1-4020-8764-6} (2008)

\bibitem{Lauritsen}
	K.B. Lauritsen, H.C. Fogedby,
%	\tarttitle{Critical exponents from power spectra}
	in {Journal of Statistical Physics} 72, 189:205 (1993)

\bibitem{Manna}
	S.S. Manna
%	\tarttitle{Two-state model of self-organized criticality}
	in {Journal of Physics. A: Mathematical and General} 24, L363:L369 (1991)
	
\bibitem{Oslo_Model}
	 K. Christensen, A. Corral, V. Frette, J. Feder, and T. J¿ssang,
%	\tarttitle{Critical exponents from power spectra}
	in {Phys. Rev. Lett} 77, 107:110 (1996)
	
\end{thebibliography}
\end{document}